\documentclass[preprint]{revtex4}

\usepackage{rotating} 
\usepackage{graphicx}
\usepackage{graphics} 
\usepackage{url}      
\usepackage{color}

\makeatletter
\def\url@leostyle{%
  \@ifundefined{selectfont}{\def\UrlFont{\sf}}{\def\UrlFont{\small\bf\ttfamily}}}
\makeatother
\usepackage[pdftex]{hyperref}
\hypersetup{
pdftitle={Content and Network Dynamics Behind Egyptian Political Polarization on Twitter},
pdfauthor={J. Borge-Holthoefer, W. Magdy, K. Darwish, I. Weber},
pdfkeywords={Egypt, Twitter, mobilization, polarization, opinion switch},
bookmarksnumbered,
pdfstartview={FitH},
colorlinks,
citecolor=black,
filecolor=black,
linkcolor=black,
urlcolor=black,
breaklinks=true,
}

\begin{document}

\title{Content and Network Dynamics Behind Egyptian Political Polarization on Twitter\footnote{{\em To appear in the} Proceedings of the 18th Conference on Computer-Supported Cooperative Work and Social Computing CSCW (2015)}}

%
%

\author{Javier Borge-Holthoefer}
\email{jborge@qf.org.qa}
\affiliation{Qatar Computing Research Institute. Doha, Qatar}

\author{Walid Magdy}
\email{wmagdy@qf.org.qa}
\affiliation{Qatar Computing Research Institute. Doha, Qatar}

\author{Kareem Darwish}
\email{kdarwish@qf.org.qa}
\affiliation{Qatar Computing Research Institute. Doha, Qatar}

\author{Ingmar Weber}
\email{iweber@qf.org.qa}
\affiliation{Qatar Computing Research Institute. Doha, Qatar}

\begin{abstract}
There is little doubt about whether social networks play a role in modern protests. This agreement has triggered an entire research avenue, in which social structure and content analysis have been central --but are typically exploited separately.

Here, we combine these two approaches to shed light on the opinion evolution dynamics in Egypt during the summer of 2013 along two axes (Islamist/Secularist, pro/anti-military intervention). We intend to find traces of opinion changes in Egypt's population, paralleling those in the international community --which oscillated from sympathetic to condemnatory as civil clashes grew. We find little evidence of people ``switching'' sides, along with clear changes in volume in both pro- and anti-military camps.

Our work contributes new insights into the dynamics of large protest movements, specially in the aftermath of the main events --rather unattended previously. It questions the standard narrative concerning a simplistic mapping between Secularist/pro-military and Islamist/anti-military. Finally, our conclusions provide empirical validation to sociological models regarding the behavior of individuals in conflictive contexts.
\end{abstract}


\maketitle

\section{Introduction}
When Time magazine nominated the anonymous protester as the Person of the Year in 2011, it was already clear that online technologies occupied (and occupy) a central spot as the backbone of modern civil protests. Beyond that cycle of turmoil, subsequent events have only reinforced such impression. Public availability and down-to-the-second temporally resolved data have given Twitter an incomparable advantage in addressing important questions, for instance how social networks actually facilitate the emergence and diffusion of protests \cite{gonzalez2013broadcasters}, or how their contents change in time.

This is of particular importance in the Middle-East and North Africa (MENA) region, where chronic political unrest is combined with a speeded adoption of online technologies. Furthermore, the rapid growth of social networking sites since 2011 shifts away from typical social and entertainment uses of online media, towards those that are more political and civic \cite{arab2011}. Indeed, recent turmoil in MENA countries has provided good case studies for online interaction: Iran (2009), Tunisia, Lybia, Egypt and Bahrain (2011), Palestine and Israel (ongoing Gaza conflict) and Syria (2011-present), have witnessed social revolts, violence and even some government changes. And in all of those, online social networks have been claimed to have a key role.

Since 2011, Egypt's society is known to be highly polarized with the two dominant poles typically labeled as ``Islamist'' and ``Secularist''. On top of this, the country has experienced much political upheaval during the past year. In such heavily polarized atmosphere, the movement named ``{\em Tamarod}'' (meaning \textit{Rebellion} in Arabic) was incepted in Egypt aiming to overthrow president Morsi a few months into his presidency. The efforts of the movement culminated in mass demonstrations and counter-demonstrations on June 30, 2013 and resulted in military intervention (coup or revolution\footnote{Different political camps use different terminology to refer to the events surrounding the military intervention on July 3, 2013.}) to displace Morsi on July 3, 2013. The military takeover and ensuing opposition to it has led to significant tension and much violence.
As a result, a new pro- {\em vs.} anti-military intervention dimension has emerged. This latter axis has been widely considered in mass media to be aligned with the previous polarization, in the sense that Islamist/Secularist coincide with anti-military/pro-military, respectively. However, during the violent aftermath of the army's ascent to power, some observers speculated that some of the supporters of the military intervention might have changed their opinion, due to what is mostly viewed --including international observers-- as excessive use of force against protesters, with at least several hundred killed.

In this paper, we examine how these events have transpired on Twitter and some of the underlying phenomena.
One angle that we are particularly interested in is one of changes in volume or ``loudness'' of different political camps over time and if these were driven by opinion changes. This question is partly motivated by the fact that ``the winds turned'' several times for the Muslim Brotherhood, and foreign governments were also struggling to decide which side to support \cite{news1}.

In particular, we attempt to address some inter-related research questions which --we hypothesize-- can be tackled using large amounts of Twitter data. First, we examine whether ``opinion switching'' between both camps can be detected and estimated. If so, we can move on and question further whether perceived opinion changes are due to (i) people actually switching sides, or (ii) the relevant camps becoming increasingly louder or more muted.

In quantifying the answer to these questions, we learn about the underlying psychological and sociological mechanisms which are at play in conflictive contexts. For instance, it might be more benefitial --or less costly-- for an individual to withdraw from an ongoing conflict than to actually switch sides: with a similarly minded social neighborhood, an actor is faced with the ``volunteer's dilemma'' (in this case, who switches side first); from which the rational outcome is to become mute~\cite{chenoweth2011civil}.

Beyond partial accounts, we approach these questions both from the structural (network) side and the semantic (content) side.
Concerning content, we used a set of manually labeled hashtags to build a pro- or anti-military-intervention classifier, which is used to classify tweets based on its textual content. On the network side, we used retweets of hand-labeled seed users to derive a Secular {\em vs.} Islamist leaning for users.
We apply our methodology on a set of nearly 6 million Arabic tweets crawled between June 21 and September 30, 2013. From these, we reconstructed the network of Twitter users who authored tweets in the collection, and we crawled all meta information for 120,000 users along with their latest 3,200 tweets prior to December 2013. All the tweets and Twitter user IDs in our collection are publicly available\footnote{\url{http://alt.qcri.org/~wmagdy/EgyMI.htm} Only IDs are provided in order not to violate Twitter's terms and conditions.}, so that researchers can replicate the exact dataset we used for possible future studies.

We are well aware that studies conducted solely on Twitter data have certain limitations and drawbacks. From a general point of view, Twitter can not be regarded as {\em the} voice of civil society. As a matter of fact, Twitter (and online social networks in general) have been adopted by a minority in these societies. Furthermore, the adoption of these new technologies are not uniformly spread, which implies that the data are, most probably, biased. And yet, we are confident that our study stands on a reliable ground. Though limited, Twitter is a valuable monitor of social dynamics and it is doubtless an influencing actor in modern societies, partially shaping the flow of information among individuals and, in many occasions, performing as an agenda-setting actor. Methodologically, our longitudinal approach tracks not absolute values (raw signal), but rather relative changes (first derivative of the signal) which diminish the impact of inherent biases. Whereas biases are expected to strongly impact the absolute level of, say, Morsi support on Twitter, trends and the direction of \emph{changes} in this level are expected to be more robust.

We find that despite --or because of-- the dramatic and violent events there is very little evidence of users changing sides or ``switching''. We look at switching between Secularist and Islamist camps and between pro-military and anti-military camps.  Our network and content analyses indicate that less than 5\% of users switched sides.  Instead, the main narrative seems to be one of pro-military intervention and Secular users being dominant in terms of volume leading up to July 3, and anti-military intervention and Islamist users gaining in volume afterwards. Furthermore, in contradiction to the dominating narrative in mass media, the correlation between being a secular and a supporter of military intervention is far from perfect. However, some correlation was noticed between being an Islamist and against the military intervention. 

\section{Egypt's {\em coup d'etat}: timeline}
\label{background}
In this section, we summarize the major events that unfolded in Egypt during the summer of 2013. In general, we can distinguish three major periods: the lead up to the June 30 protests; the military intervention on July 3; and the aftermath of the intervention, including the violent crackdown against the Rabia sit-in. 
The chronology should help in understanding the conflict in the time period that we cover. 
We also invite the reader to have a look at a Wikipedia article on the topic\footnote{\url{http://en.wikipedia.org/wiki/2013_Egyptian_coup_d\%27\%C3\%A9tat}} and an Al-Jazeera interactive timeline tool for the Egyptian turmoil\footnote{\url{http://www.aljazeera.com/indepth/interactive/2013/08/2013817122637981237.html}} for more information.

The major events were as follows:
\begin{itemize}
\item \emph{{\bf Late April, 2013:}} A movement named {\em Tamarod} started in an effort to force the Egyptian president Mohamed Morsi out of office. The movement, which called for mass protests on June 30, 2013, claimed to have collected 22 million signatures by June 29. It was aided by strong media support and major power and gas crises~\footnote{\url{http://www.spiegel.de/international/world/egyptian-army-gives-morsi-of-muslim-brotherhood-a-48-hour-ultimatum-a-908823.html}}. 
Tamarod is often branded as a Secularist or pro-military movement.
\item \emph{{\bf June 28 and 30:}} Large anti-Morsi protests take place in Cairo and across the country. Counter-demonstrations in support of Morsi also begin with major sit-ins in Rabia and Nahda squares. The sit-ins continue non-stop until they are disbanded on August 14. Morsi supporters are often branded as Islamists.
\item \emph{{\bf July 1:}} The military issues a two-day ultimatum to the Egyptian president to accept the demands of Tamarod and to call for early presidential elections.
\item \emph{{\bf July 2:}} The president, in an address to the nation, concedes to some demands from Tamarod. However, he refuses to step down or to call for early elections~\footnote{\url{http://english.ahram.org.eg/NewsContent/1/64/75538/Egypt/Politics-/Egypts-Morsi-defies-calls-to-step-down,-offers-opp.aspx}}.  
\item \emph{{\bf July 3:}} General Sisi, the head of the military, orchestrates a coup to overthrow the president, suspend the constitution, disband the elected legislature, arrest Morsi's most prominent supporters, and shutdown pro-Islamist television stations. A new interim president is appointed. Protests in support of Morsi continue and grow.
\item \emph{{\bf July 8:}} Clashes between security forces and anti-coup protestors erupt in front of a National Guard compound leading to many deaths.
\item \emph{{\bf July 24:}} General Sisi asks his supporters to demonstrate, so as to obtain a ``mandate'' to fight ``possible terrorism''.
\item \emph{{\bf July 26:}} Pro-military protests take place in Tahrir square in response to General Sisi's demands.
\item \emph{{\bf July 27:}} Clashes between security forces and anti-coup protestors erupt near Rabia square, leading to the death of dozens (mostly demonstrators).
\item \emph{{\bf August 14:}} Security forces crack-down on the sit-ins, leading to the death of hundreds of anti-coup protesters. Demonstrators attempt to establish new sit-ins, but these are as well violently disbanded.
\item \emph{{\bf August 16:}} Major anti-coup protests take place, leading again to the death and arrest of many protesters.
\end{itemize}

Besides the previous specific dates, protests and clashes are intensified on Fridays (prayer day for Muslims) with continuing sit-ins in Rabia and Nahda squares up until August 14.

\section{Related Work}\label{sec:related}
For a helpful contextualization of our work, we focus on three research avenues. First, a general framework is that of temporally resolved social dynamics, i.e.\ works in which scholars attempt to track the evolution of social phenomena (as opposed to a static snapshot). Within this area, we highlight the studies that zoom in on political events: the growth of grassroots movements, recruitment processes and (bi)polarization of opinions. A second major area --indeed with much work devoted to it-- is that of the Arab world, digital media and/or the political/civic use of it. Third, we briefly discuss sociological accounts of opinion dynamics --``switches'' and silencing.

In the first line of research, it is possible to identify some works which provide longitudinal accounts of political (protest) events. Some examples are the work by Borge and collaborators \cite{borge2011structural} (devoted to the growth of the ``Indignados'' movement in Spain) or Gonz\'{a}lez-Bail\'{o}n {\em et al.} \cite{gonzalez2011dynamics} (recruitment processes again in the 2011 protests in Spain). But closer to the current approach, we pay attention to Weber {\em et al.}'s work \cite{webergarimella13asonam,weberetal13asonam}, who explore the bipolar political scenario in Egypt (Secular {\em vs.} Islamist). To do so, they track the retweeting behavior from certain seed users as a signal to deduce a user's political orientation. Furthermore, they transfer the polarity from users to hashtags based on usage patterns --such that a hashtag is tagged as Islamist, Secular or ``neutral''. In doing so, the authors provide a means to assess real-time online polarization. They provide anecdotal evidence that the increases in this polarity score anticipate periods of political violence.

Needless to say, polarization has been a focus of interest before. Starting with the NOMINATE score~\cite{poolerosenthal85ajps}, to quantify the statement that U.S. politics follows a 1-dimensional left-to-right schema, much research has followed. Focusing on digital media and U.S.-centered politics (i.e.\ left-{\em vs.}-right polarization), Adamic {\em et al.} \cite{adamicglance08linkkdd} and Conover {\em et al.} \cite{conover2011political} capitalized on data from the blogosphere and Twitter respectively. The time-resolved retweet graphs constructed from our data set present some similarities to Adamic's work; and yet, the temporal dimension is missing in these articles.

Second, there is a large body of work on the so-called Arab Spring, some of which zoom in on the Egyptian Revolution in particular. These depart from the present work insofar they typically include some discussion on the role of social media during the protests and the revolution, but do not use data from social networks (nor Twitter in particular) to approach polarization or major events. A notable exception is work by Mostak \cite{mostak2012}, who tests the political hypotheses that ``Islamism in an ideology of the poor'' with online data. To approach the question, he looks for geographical correlations between census information for income and house value/size, and estimates for how much Islamist activity on Twitter there is originating from a particular region. Down to conflictive situations, we find a plethora of works devoted to the Arab Spring --as the expression ``Twitter Revolution'' was coined and settled in mass media: some examples can be found in \cite{lim12jc,starbirdpalen12cscw,attiaetal11ecra,ohetal12icis,khamis11ijoc,azab12ijep}, often from a qualitative point of view or relying on surveys. With the use of data from blogs, Al-Ani {\em et al.} \cite{al2012egyptian} explore alternative news sources --beyond the government-supplied versions of events. However, none of them quantifies online polarization, or provide a longitudinal point of view. Other countries and conflicts have of course caught the attention of researchers, such as Tunisia \cite{wulf2013ground} and Palestine \cite{wulf2013fighting}. Closer to this work, Choudhary {\em et al.} \cite{choudharyetal12cacm} performed time-resolved sentiment and response analysis to Twitter activity during the events in 2011, that eventually led to the displacement of Mubarak and the onset of a transitional period. Apparently, they used English tweets coming from Egypt and relevant news sources. In our work, we focus on Arabic tweets --clearly underrepresented in scholarly articles-- to better model people in the region.

Finally, we find literature that focuses on the monitoring of opinion changes and the existence of online ``silent crowds''. Far from the political debate, some researchers have tried to tackle the fundamental differences between the ``vocal minority'' and the ``silent majority'' \cite{mustafaraj2011vocal}. In this work, a {\em caveat} is placed to consider the significant differences between those highly active online media users --who dominate the ongoing discussions over social networks-- and the vast majority of people who participate much less frequently. The same line of argumentation is found in Venkataraman {\em et al.} \cite{venkataraman2012measuring}. A step closer to our politically motivated work, Lin {\em et al.} \cite{lin2013voices} exploit sentiment analysis to track political opinions in U.S. --although their method is intended for general purposes. Common to our framework, a real-time, evolving opinion-shift account is offered. Finally, on a more theoretical ground, the work by Chenoweth {\em et al.} \cite{chenoweth2011civil} suggests that there are indications that, in a conflictive environment, members of opposing groups do not necessarily switch sides to tip the balance in favor of one of the groups, but they may merely withdraw their support.

\section{Data Collection}
\subsection {Collecting Relevant Egyptian Tweets}
We collected Arabic tweets in the period between June 21, 2013 and September 30, 2013. This period covers the major events in the aforementioned chronology.
We used Twitter4J APIs\footnote{https://code.google.com/p/tweet4j/} to collect the tweets matching the query ``\textit{lang:ar}'', (so as to retrieve Arabic tweets).
On average, the number of collected tweets per day was 3 million. The Arabic text of the collected tweets was pre-processed using state-of-the-art normalization technique for social Arabic text \cite{darwish2012language} to facilitate later filtering. The normalization includes letter normalization, diacritics removal, decorative text replacement and word elongation compression. 

To extract the tweets that are relevant to Egypt, we constructed a rich set of Boolean queries that cover different facets of Egyptian politics. The set contained 112 requests that represent key players in the Egyptian political sphere: government representatives, political parties, opposition leaders, media anchors and the Twitter accounts of some relevant players.
Each Boolean request is prepared to handle different spellings of names, acronyms, and nick-names used by different political groups. As such, an enriched set of queries underlies every request in order to retrieve relevant tweets from the Arabic news platform TweetMogaz\footnote{\url{http://www.tweetmogaz.com}} \cite{magdy2013tweetmogaz,magdy2014tweetmogaz}. TweetMogaz automatically generates news about Egypt, taking Twitter as the source. By default, the platform uses 84 queries and applies an adaptive tracking algorithm to enrich them, relying on ongoing news \cite{magdy2014adaptive}. For this work, we manually revised the automatically enriched keywords to ensure a high precision and recall.
On top of that, we also considered context keywords, i.e.\ relevant to that particular period: ``revolution'', ``June 30'', ``coup'', ``Rabia'' and ``massacre''. This extra set was picked by two native Egyptian, who regarded these words as most meaningful to the studied turmoil episode.

\begin{figure*}
\centering
\includegraphics[width=\textwidth]{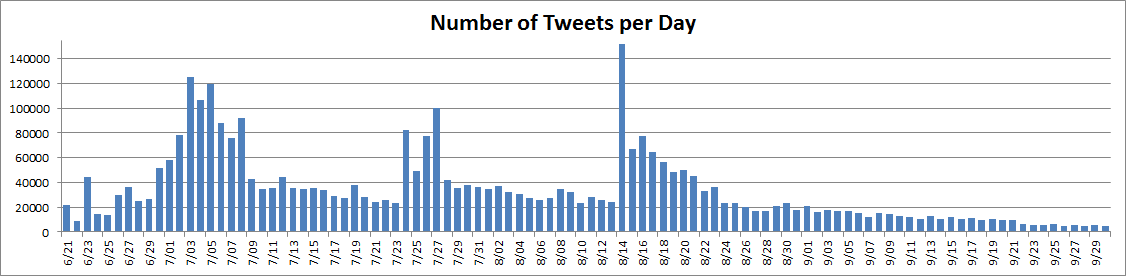}
\caption{Distribution of the collected tweets about Egypt over time}
\label{TweetsVolume}
\end{figure*}

The number of extracted tweets matching those queries during the mentioned time window was 5.9 million. To ascertain the accuracy of the matching items, we manually judged a set of 500 random tweets, from which only 5 were assessed as not relevant, rendering a 99\% estimated accuracy.
Figure~\ref{TweetsVolume} shows the volume of tweets collected on the filtered set over time. As expected, major events led to peaks in the number of tweets on the days they occurred. There is clear mapping between the peaks in Figure~\ref{TweetsVolume} and the events listed in the timeline section (see above), which hints at the fact that Twitter is --among others-- a faithful and sensitive monitor of real-world (offline) life.

\subsection{Collecting Users' Network Data}
We identified Twitter accounts that authored/retweeted more than 10 tweets in our collection, to find out their network information for network analysis. In doing so, we guarantee that the monitored users did engage, at some time or another, the ongoing crisis in Egypt. This led to the identification of nearly 120,000 Twitter accounts, out of roughly 1 million in our collection of Egyptian tweets. For each user, we extracted their network information, which comprehends the number of followers, friends and, most importantly, the last 3,200 tweets they authored or retweeted. These latest tweets allowed us to measure the network dynamics of each user through their activities including retweets, mentions, or replies to other accounts. The resulting information exceeded 750GB of user data and more than 300 million tweets for these users. 
Table~\ref{datasummary} summarizes the obtained tweets data collection.

\begin{table*}[width=6in]
  \centering
	\begin{tabular}{ | l | p{4cm} | p{5.5cm} | p{4.5cm} |  }
		\hline
		Data & Description & Included Information & Size \\
		\hline
		Egyptian tweets & Egyptian tweets between June 21 and October 1, 2013 & Tweet ID, text, Twitter user, and Timestamp & 5,902,086 tweets \\
		\hline
		User Data & Profiles of Twitter users with more than 10 tweets in the Egyptian tweets set. & Username, screen name, ID, 
 number of followers and followees.\linebreak List of last 3,200 posted tweets by the account (prior to the mid of December 2013) & 121,003 profiles. \linebreak Recent 3,200 tweets per profile \\
		\hline
	\end{tabular}
	\caption{Tweets data collection used in our study}
	\label{datasummary}
\end{table*}

\section{Content-based Analysis of Tweets}
In this section, we describe the analysis we conducted on tweets, based on their textual content. To this end, we built a supervised classifier that groups tweets according to their content, i.e.\ whether they express support (pro-MI: pro-military intervention) or opposition (anti-MI) to the military intervention, plus a ``neutral'' category as well. We use this classifier to measure the response to the events in the time window of interest. Later, we deepen our analysis to track for possible opinion changes at the individual level.

\subsection{Classifying Tweets as Pro/Anti-Military Intervention}
The design of the classifier comprehends three main steps. First, we constructed lists of seed hashtags that are either exclusively or predominantly used by either of the two groups. Then, we expanded those lists to identify more tweets that could potentially belong to either group. Finally, we employed a supervised method to classify the tweets. 

For the first step (seed lists), we extracted the most frequent hashtags in our collection, which were  then tagged as either most likely pro-MI or anti-MI group. The rest of hashtags were neglected. Briefly, we assumed pro-MI hashtags to express support for the army, calls for a regime change or a {\em revolution} (as opposed to {\em coup}), opposition to Morsi or his regime or attacks against the Muslim brotherhood. Conversely, the group of hashtags categorized as anti-MI expressed maligning the members of the Supreme Council of Armed Forces (SCAF), the usage of {\em military coup} (as opposed to {\em revolution}), expression of support for Morsi or his regime, or rejection to the crackdown against the sit-ins. The lists contained the highest frequent 150 hashtags for each polarity, see Table~\ref{HashtagExamples} for some examples.  

With the seed lists at hand, we heuristically classified tweets in our collection as belonging to the pro-MI or anti-MI groups if they exclusively contained hashtags belonging to the corresponding lists, for either of the groups. If a tweet contained no hashtags indicating polarity or contained hashtags from both lists, it was discarded. Using the filtered tweets, we identified new words and hashtags that exclusively appeared in either group, but not in both. We recursively added these words and hashtags to their respective lists. We repeated this expansion step four times. The resultant lists for the pro-MI or anti-MI group lists contained 1,105 and 1,487 words and hashtags respectively. Exploiting these final filtering lists, we constructed a subset of the tweet collection containing 12\% of the tweets in the whole dataset.

Finally, from the filtered tweets subset we randomly selected 500 tweets containing hashtags from each list. These 1,000 tweets were manually verified and judged as exclusively belonging to pro-MI, anti-MI or neutral. Judgments were performed by two Egyptian Arabic speakers, who are keenly aware of the situation in Egypt. The inter-annotator agreement between them was above 93\%. In case of disagreement, the annotators discussed over the tweets until consensus was reached. After this process, the agreement exceeded 96\%. For the remaining tweets (33), one of the judgments was selected at random. Table~\ref{LabelDistribution} shows the distribution of the 1,000 tweets over the three groups.  We used a multi-class Support Vector Machine (SVM) classifier to classify tweets as pro-MI, anti-MI, or neutral. We employed the SVMLight Multiclass implementation~\cite{crammer2002algorithmic} for classification. We used the following features: word unigrams, word bigrams, and hashtags. We performed 20 fold cross validation over our 1,000 labeled tweets where each fold (containing 5\% of the examples) was used for testing and the 19 other folds (containing 95\% of the examples) were used for training. The average accuracy for the folds was 87.0\% with a standard deviation of 3.4\%. Table~\ref{LabelConfusion} shows the classifier confusion between the guessed and actual tags.

\begin{table}
\centering
\includegraphics[width=0.75\columnwidth]{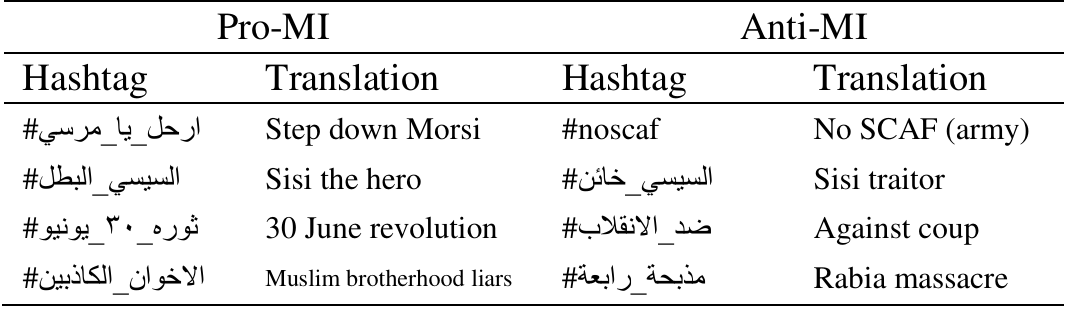}
\caption{Examples of seeding hashtags used for Pro/Anti-military classifier}
\label{HashtagExamples}
\end{table}

\begin{table}[h]
  \centering
	\begin{tabular}{ |c|c|c|}
		\hline
		Pro-MI	& Neutral	& Anti-MI\\\hline
		35\%	& 9\% &	56\%\\\hline	
	\end{tabular}
	\caption{Distribution of tags in training/test set}
	\label{LabelDistribution}
\end{table}

\begin{table}[h]
  \centering
	\begin{tabular}{|l|c|c|c|}
		\hline & \multicolumn{3}{c|}{guess}\\
		\hline
		Truth & Pro-MI	& Neutral	& Anti-MI\\\hline
		Pro-MI &	\textbf{0.817} &	0.064 & 0.119\\
		Neutral	& 0.125	& \textbf{0.736} &	0.139\\
		Anti-MI &	0.054 & 0.022	& \textbf{0.924}\\\hline
	\end{tabular}
	\caption{Confusion between truth and guessed labels for classes}
	\label{LabelConfusion}
\end{table}

\begin{figure}[ht]
\centering
\includegraphics[width=\columnwidth]{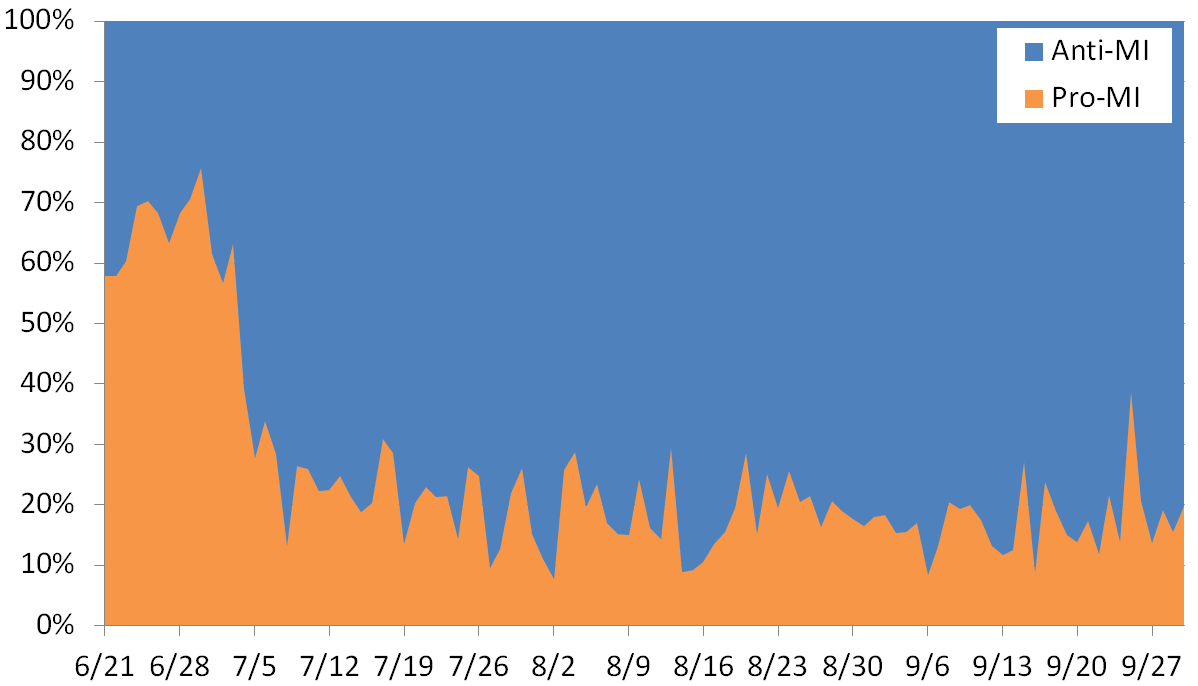}
\caption{Proportion of tweets per day according to polarity of supporting/opposing military intervention}
\label{PolarityVolume}
\end{figure}

Though the undetermined tweets may contain pro-MI or anti-MI tweets, our interest focuses on finding the distribution of tweets belonging to both groups over time. Thus we favored precision. Figure~\ref{PolarityVolume} plots the proportion of tweets that were classified as opposing (anti-MI: blue) or supporting (pro-MI: orange) the military intervention over time. As illustrated in Figure~\ref{PolarityVolume}, before the military intervention on July 3 the proportion of tweets belonging to the pro-MI group was dominant (55\% to 75\%). After July 3, pro-MI tweets witnessed a significant drop, while anti-MI tweets rose substantially. This can be attributed to many different reasons that we attempt to analyze in the next section.

\begin{figure*}
\centering
\includegraphics[width=\textwidth]{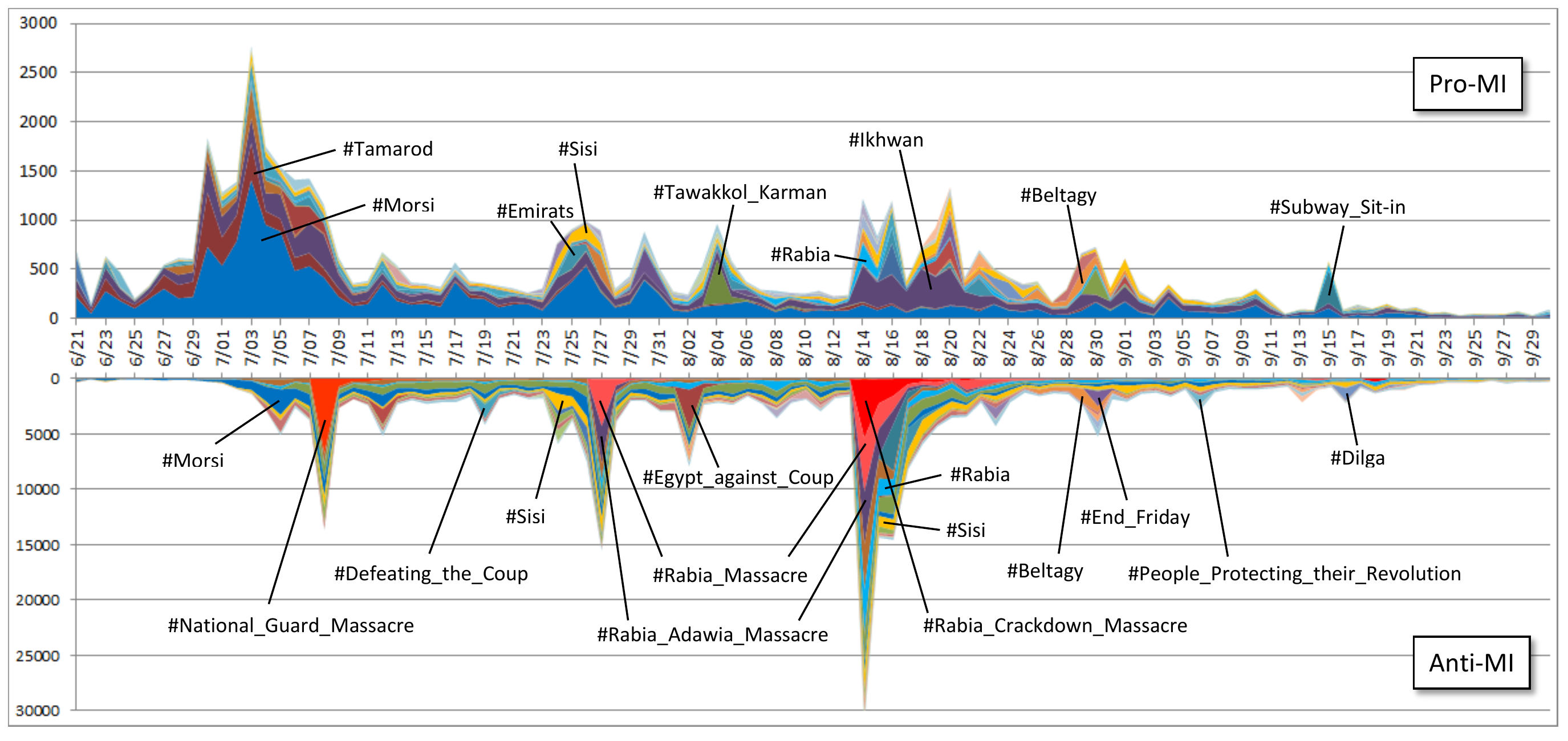}
\caption{Number of occurrences of the most outstanding 30 hashtags appeared in the pro-MI (upper) and anti-MI (lower) tweets between 21 June and 30 Sep. 2013. Note the scale for the anti-MI plot, which is 10 times the scale of the pro-MI graph. Presented hashtags are translations of the original Arabic ones.}
\label{tweet-dist}
\end{figure*}

Given the content of our filtered collection, we analyze the top used hashtags by each group over time. We do so in order to understand the topics of interest for each of the groups and how they evolved in response to major events. We identified the 30 hashtags that achieved the highest per-day volumes for any of the days in the full period, for both groups. We eliminated hashtags, such as \textit{\#Egypt}, that were mentioned everyday, achieving the highest aggregate volume over all the days --but never exhibited a burst in volume for any given day. Using this scheme, a hashtag that is mentioned 10,000 times in one day is more important than another that is mentioned 1,000 times daily over a span of 100 days. A hashtag showing a spike in volume would likely indicate changes in topics or interest for groups.
Figure~\ref{tweet-dist} shows the number of occurrences of the identified 30 hashtags in the period of study. The upper panel shows the most frequent hashtags on the pro-MI group, while the lower panel is devoted to the most frequent ones on the anti-MI group. Surprisingly, the overlap between hashtags representing the topics discussed by each group is almost nonexistent. Only four (out of 30) hashtags are in common in the two groups (\textit{\#Morsi}, \textit{\#Sisi}, \textit{\#Rabia}, and \textsl{\#Beltagy}). However, the use of these hashtags were different in scale, time periods, and sentiment.

Upon inspection of the tweets and hashtags of the pro-MI group, they are focused more on individuals and organizations. The top hashtag is \textit{\#Morsi}. The use of the tag peaked around his ousting on July 3, and the relative volume of the hashtag for the rest of the period continued to be high. Similarly, tweets critically mentioning \textit{\#Ikhwan} (Muslim Brotherhood) persisted with relatively high volume for the whole period. \textit{\#Tamarod} peaked around Moris's deposition, but dwindled quickly afterwards. Other pro-MI tweets belonging to persons or organizations included support for \textit{\#Sisi} (the head of the military), attacks directed at \textit{\#Tawakkol\_Karman} (Yemeni Nobel prize laureate who was critical of the military takeover) or joy at the arrest of \textit{\#Beltagy} (former parliamentarian and a prominent anti-MI figure). Other prominent hashtags can be found related to criticism towards the sit-ins in \textit{\#Rabia}, with a peak on August 14, and of the \textit{\#subway\_sit-in}, with a peak on September 15.

Anti-MI tweets and hashtags instead focused mostly on events, particularly those involving violence and loss of life (\textit{\#National\_Guard\_Massacre}, July 8), \textit{\#Rabia\_Massacre}, \textit{\#Rabia}, \textit{\#Rabia\_Adawia\_Massacre}, \textit{\#Rabia\_Crackdown\_Massacre} (July 27 and August 14) and \textit{\#Dilga} (September 16). They also include hashtags for announced protests on specific days or weeks, with activities against the military (\textit{\#Defeating\_the\_Coup}  on July 8; \textit{\#Egypt\_against\_Coup}, August 2; \textit{\#Last\_Friday}, August 8; or \textit{\#People\_Protecting\_their\_Revolution}, September 6). The most mentioned individuals in tweets where \textit{\#Morsi}, mostly around the time of his ouster, general \textit{\#Sisi}, distributed over the entire period, and \textit{\#Beltagy}, around the time of his arrest.

\subsection{Detecting Users Polarity Switching}
With a reliable classifier at hand, which provides content-based polarity classification, we can observe the changes in polarity at the Twitter user-level over time. A user polarity switch means that a user who was tweeting pro-MI tweets at $t_{0}$ moved to authoring anti-MI tweets at $t_{1} > t_{0}$, or viceversa. We apply this analysis to understand whether this was the reason behind the large decline in the number of pro-MI tweets after the military intervention on July 3 as shown in Figure~\ref{PolarityVolume} and the increase in anti-MI tweets. 

Two hypotheses are compatible with the observed pattern changes, namely: people switched camps in response to events; or, simpler, one camp became quieter while the other becoming louder. We envisage that the latter is the mechanism behind the phenomenon --in accordance with previous literature \cite{chenoweth2011civil}. And yet empirical evidence is lacking in a violent context, thus the relevance of the present work.

Detecting switch in user's political leaning requires a minimum number of tweets of that user over the studied period to monitor his leaning over time. The confidence of detecting a switch in leaning increases the larger the number of pro/anti-MI tweets are examined for a given user: more tweets supply more evidence of the user's leaning. Unfortunately, most of the users in our collection did not have large number of tweets classified as pro/anti-MI. Therefore, we relied on a small portion of users who had at least $n$ classified tweets, $n \in \{5, 10, 15, 20\}$. Examining leaning switch of users based on only 5 tweets might be inaccurate, however this leads to the examination of more users in our test set; whereas examining users with at least 20 tweets in our collection leads to a smaller test set, though yielding a higher confidence.

\begin{figure}
\centering
\includegraphics[width=\columnwidth]{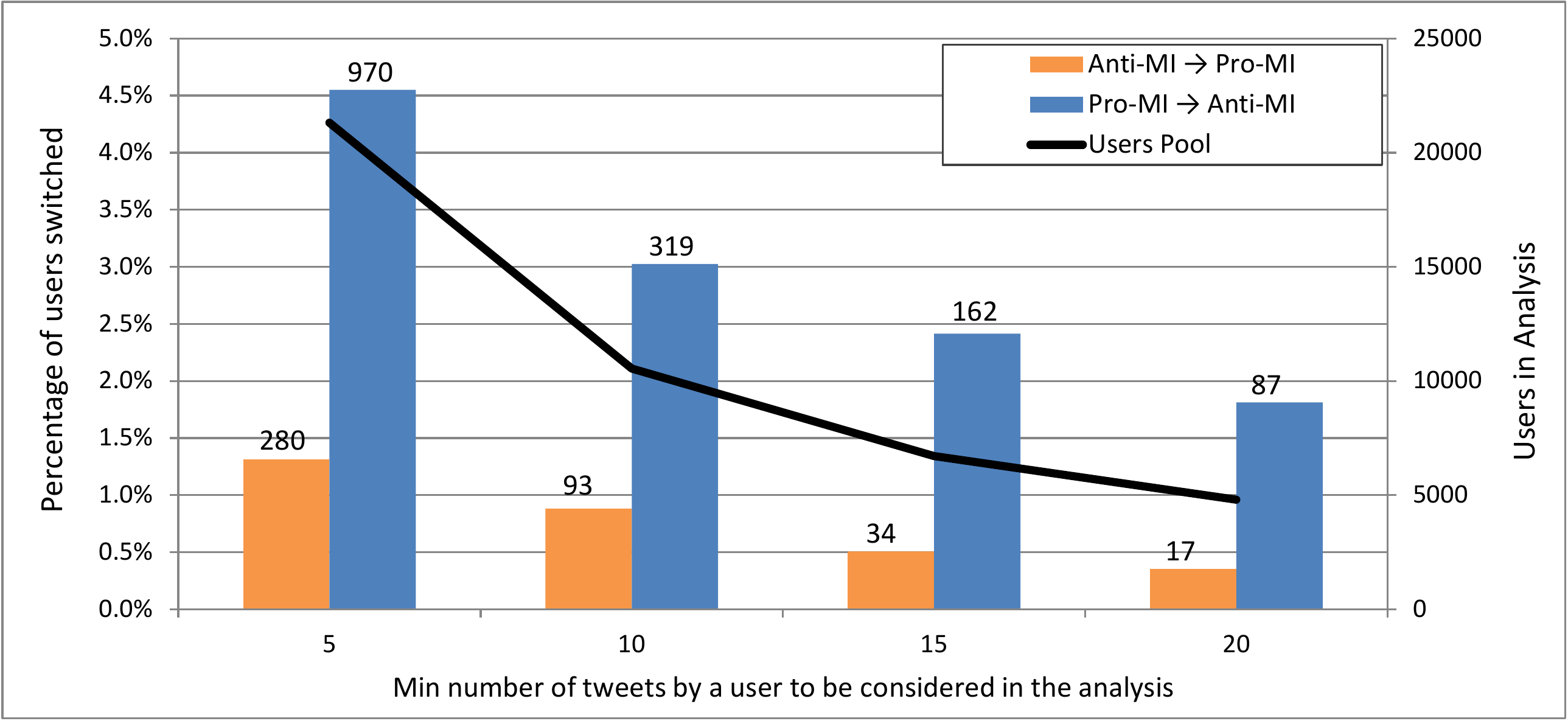}
\caption{Percentage of users switched from one polarity to the other, and the number of users used for the analysis of switching for different values of \textit{n}}
\label{PolaritySwitch}
\end{figure}

To consider a user polarity switch, we examined the first and last third of tweets for the user. If the average polarity changed from the first third compared to the last third, we assume that the user swapped her leaning. Otherwise, we consider that the user held a coherent (stable) opinion over time. Figure~\ref{PolaritySwitch} reports the percentage of used switched from anti-MI to pro-MI, or vice versa, and presents the number of users examined in the analysis for different values of $n$. As shown, for all $n$ values, which reflects different confidence levels, the number of user switching sides are minimal, where it did not exceed 5\% of the total number of users examined. For example, the largest percentage when $n = 5$, out of 21,312 users examined, only 970 switched from pro-MI to anti-MI, and 280 switched from anti-MI to pro-MI. The percentage of switching is even less for larger $n$ values, indicating that the higher percentages for smaller $n$ might causes by random noise in the classifier. In all the switches, the number of users switched against the military intervention are at least triple the number of users switched to support the military intervention.

Our ``first third, last third'' schema is admittedly an arbitrary one. There are, however, at least two good reasons to proceed in this way. First, an opinion change might not occur in a sudden way, but rather progressively (as events unfold). Thus, it is necessary to focus the attention on the initial opinion state and the final one, with some unobserved time in between. In doing so, we avoid the presence of noise that might ``contaminate'' our output. For this reason, a $\frac{1}{2}-\frac{1}{2}$ schema would not yield a reliable result (despite its providing larger statistics). In the opposite extreme, a more refined method would be to consider a small initial and final tweeting period (first $\frac{1}{10}$th, last $\frac{1}{10}$th, for instance). We would have obtained a more robust result, at the expense of poor statistics.

Table ~\ref{SampleSwitches} presents some example tweets (translated from Arabic) for users who switched their polarity. Example 1 in the table is for a Twitter user who was calling for protests against Morsi on June 21, and then he was calling for participation in a march against what he called the military coup on July 19. Similarly, the second example is from pro-MI to anti-MI. On July 4, he expressed approval for the overthrow of Morsi, but then on July 24, he attacked General Sisi\footnote{Sisi also means pony in Arabic.} for his request for a ``mandate''. The final example is for a user who switched from anti-MI to pro-MI, where the user showed some solidarity for protestors against military intervention, then after two and half months he showed support for Sisi and opposition to Morsi.

Although the detected user switches are interesting, they can overall be considered anecdotic: the majority of users tend to express stable opinions. Thus, it can be concluded that the significant changes in polarity proportions in Figure~\ref{PolarityVolume} are due to supporters of military intervention becoming quieter and the opponents becoming louder.

\begin{sidewaystable*}
  \centering
	\small
	\begin{tabular}{ l l p{15cm}  }
		\hline
		\textbf{Switch} & \textbf{Date} & \textbf{Tweets} (\textit{English translation}) \\
		\hline
		\emph{Pro-MI} $\rightarrow$ \emph{Anti-MI}  & June-21 & We will continue to revolt till we reach freedom. Gathering revolution from Alexandria to Cairo to oust Morsi, the sheep. \\
		 & July-19 & The Mohandseen march is closing the main streets till the police station \#No\_to\_military\_coup \\ \hline
		 \emph{Pro-MI} $\rightarrow$ \emph{Anti-MI} & July-04 & Everybody, Morsi is displaced. Damn brotherhood who pretend to be Muslims. Long live Egypt \\
		 & July-24 & I thought you were a Sisi (meaning: pony), but it turned out that you were a fox \#Sisi\_is\_calling\_for\_civil\_war \\ \hline
		\emph{Anti-MI} $\rightarrow$ \emph{Pro-MI} & July-15 & \#Egypt\_coup gas bombs are thrown inside the Fath mosque and people are trapped inside \#Egypt\_coup \\
		 & Sep-28 & Please share the photo of Saied Qotb when he was arrested in a box like the one general Sisi put Morsi, the sheep, inside \\ \hline
	\end{tabular}
	\caption{Sample tweets of Twitter users who switched polarity}
	\label{SampleSwitches}
\end{sidewaystable*}

\section{Network-based Analysis of Tweets}
From the data collected we built a sequence of temporally-evolving networks (one per day). In these networks, a weighted directed link $(i, j, w)$ is laid between two nodes if user $i$ retweeted a previous tweet by $j$ a number of times $w$. We used a 3-days overlapping sliding window scheme, where the network on day $t$ is obtained by aggregating 3 days of activity, i.e.\ retweets from $t-1$ to $t+1$; and the network on day $t+1$ spans from $t$ to $t+2$, and so on. This has some advantages, namely: a 3-day wide aggregation guarantees that retweeting patterns will reflect a certain stability --away from the noisy immediacy of events; also, the overlapping scheme allows for a finer tracking of ongoing events.

In general, the sequence of retweet-reconstructed networks exhibit some fairly constant global topological properties. As preliminary steps, ``silent'' nodes were removed and the (weakly) giant connected component was extracted (typically containing over 95\% of the original nodes). Density ($\sim 10^{-4}$), average degree ($2 \le \langle k \rangle_{t} \le 5, \forall t$) or clustering ($\sim 10^{-3}$) display small differences across time. The networks are thus very sparse and highly disassortative, in the sense that a few Twitter users get most of the retweeting activity, with hardly any presence of triadic closure.

\subsection{Community Sizes}
In this work, one of the main motivations was to adopt a network approach to track polarity evolution over time. In this direction, we disregard global (like the ones above) or microscopical (node-level) properties, to focus on changes at the meso-scale or group level. To do so, we applied a well-known community detection technique, namely label propagation \cite{raghavan2007near}, as implemented in the C-coded \emph{igraph} network analysis package \cite{igraph}. This algorithm has some highly desirable features. Apart from obtaining significant values in the modularity optimization process (the higher the modularity $Q$, the better the quality of the resulting partition), label propagation delivers excellent performance in terms of memory and time.

The algorithm runs with some predefined constraints. First, we use a list of seed users for whom the partisan leaning is out of any doubt, both in the Secularist and the Islamist side \cite{weberetal13asonam}. Part of the list was taken from Mostak \cite{mostak2012} with additional entries compiled in 2013 by an Egyptian expert\footnote{Due to largely differing degrees in popularity, the list of seed users boils down to Mohamed ElBaradei (@ElBaradei; Secularist camp) and Muhammad Morsi (@MuhammadMorsi; Islamist camp); the rest of the list has secondary effects on the output of the algorithm.}. Second, we limit the number of possible communities to two --since we track for a bipolar political scenario in Egypt. Admittedly, it is possible that the network structures (over time) allow for better, higher-$Q$ partitions with more than two communities, but that is beyond the scope of this work. Also, though Secularists and pro-MI group members do not necessarily align, Islamists were more likely to be in the anti-MI group than Secularists, as we show later. To provide an even more efficient running time, we fed the network at time $t$ with the labels obtained in the previous snapshot $t-1$, when possible (thousands of users enter and leave the network at each time step, in the latter case nodes are unassigned initially, and the algorithm eventually imposes the corresponding tag on it).

Under these considerations, the label propagation algorithm yields in general high modularity $Q$ values: the average $Q$-value per snapshot is $Q = 0.4340$ with very few variation, $\sigma_{Q} = 0.0150$. This implies that in terms of partition quality, communities remain fairly constant over time. To have a solid standpoint regarding the quality of this result, we have generated (and $Q$-optimized) 100 surrogate versions of the first snapshot. That particular snapshot has an actual $Q = 0.4431$, whereas its random counterparts achieve an average $Q = 0.0468$ with $\sigma_{Q} = 0.091$. This means the actual $Q$ corresponds to a $z$-score of 4.3151, i.e.\ it indeed represents a significant, far-from-random partition.

\begin{figure}[ht]
\centering
\includegraphics[width=\columnwidth,clip=0]{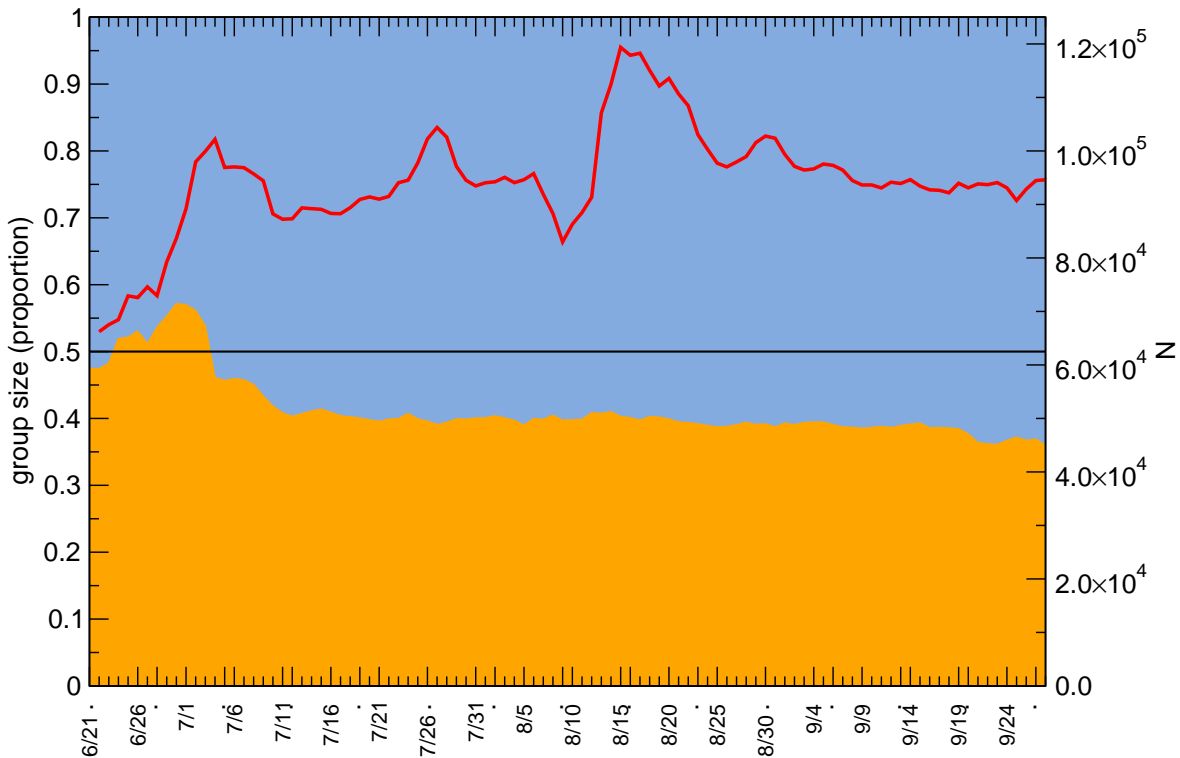}
\caption{Community sizes corresponding to Islamic (blue) and secular (orange) leanings. The red line tracks the size of the network each day.}
\label{comsize}
\end{figure}

Figure~\ref{comsize} offers a time-resolved view of community size evolution. 
During the first days of the time window --and up to the military intervention date-- Secularists outnumbered Islamists in the retweet network. On July 3, there was a crossover, and the number of Islamists remained larger until the end of the period under study. Even in bursty periods, where the network slices underwent severe changes (note for instance network sizes --red line in Figure~\ref{comsize}-- around mid-August), Secularists did not (proportionally) recruit newcomers, whereas the Islamist camp did. Unsurprisingly, the network size fluctuated over time, specially as critical events unfold.

Note that it would be overly simplistic to assume ``one Twitter user, one vote'' and that Twitter user counts perfectly match the proportion in society \cite{metaxas2011not,jungherr2012pirate,gayo2011limits}. This holds both for network-based measures as for content-based measures. Still, the trends in relative sizes of the two camps might be more robust against selection bias and other skews.

\subsection{Global Switching (Islamist vs.\ Secular)}
\begin{figure}
\centering
\includegraphics[width=\columnwidth,clip=0]{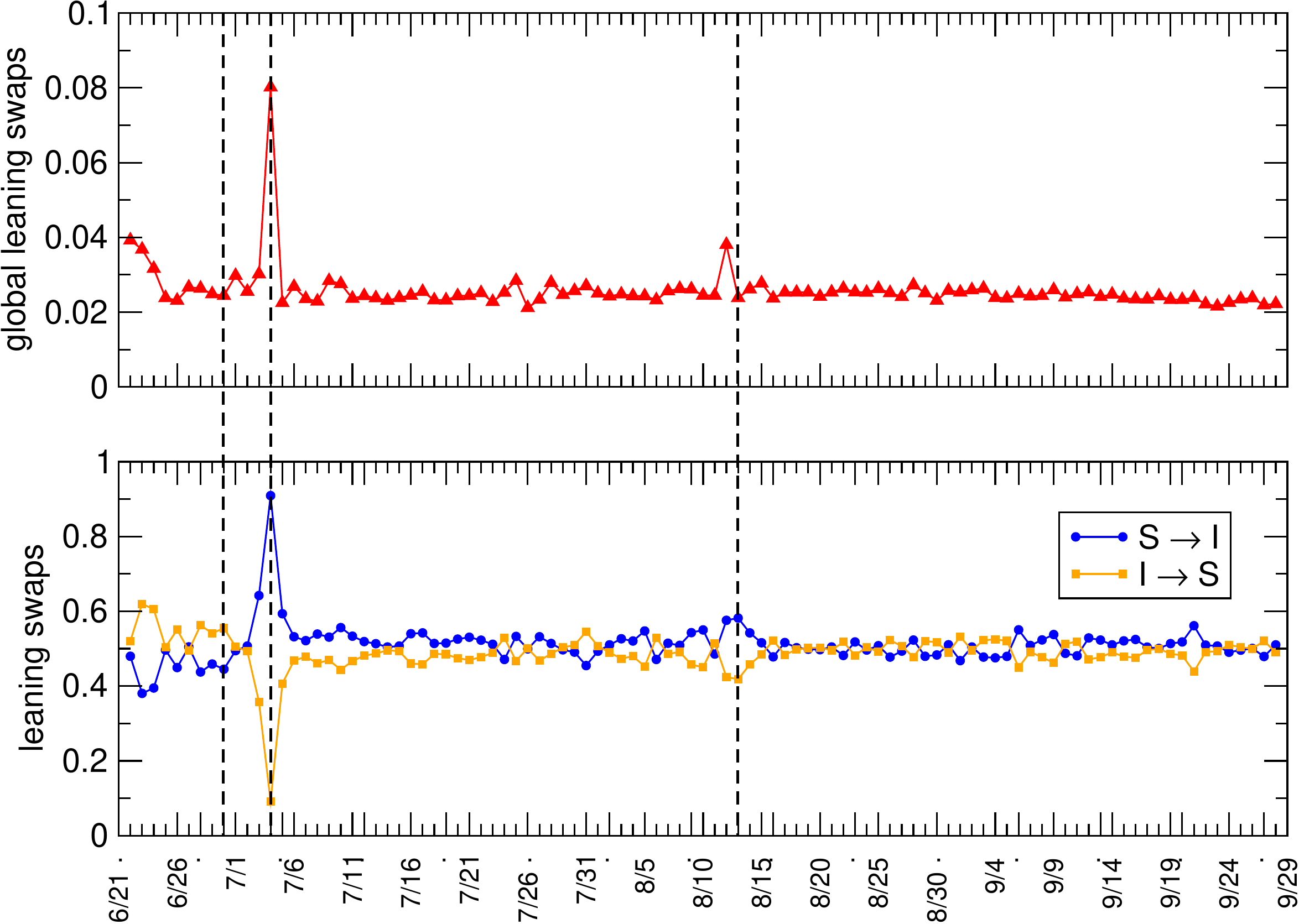}
\caption{Leaning switches occur at a very low rate (top panel; hardly 3\% of switching over time, with a few exceptions on key dates). Also, whenever swapping behavior exists, the direction of these swaps are evenly shared between parties (again, with some exceptions on key dates).}
\label{leanswap}
\end{figure}

As stated above, the Islamist dominance of the retweeting scene beyond the 3rd of July is clear. The question remains whether secular sympathizers switched to Islamist positions (as they overwhelmingly belong to the Anti-MI group), thus feeding the growth of the Islamist group as observed in Figure~\ref{comsize}. To check this, we measured the global switch ratio (Figure~\ref{leanswap}, top panel) as the number of observed label switches (from $t-1$ to $t$), over the total amount of common nodes in the networks corresponding to $t-1$ and $t$. Notably, switching remained at a very low level with a single exception (July 3) when it climbed up to 8\%. In the lower panel of the same Figure, we show the proportion of swaps in either direction, across time. Again, most of the time there is an evenly distributed amount of switching in both directions (from Secular to Islamist, and vice versa) with a noteworthy exception on July 3, when 90\% of the switches occurred from a secular to an Islamist position. Though this could be due to actual ideological turnover, it might also just point out that the focus of attention --as expressed through retweeting behavior-- is temporarily shifted towards the Islamists --whose president is ousted on that day.

\subsection{Leaning histogram vs.\ Tweeting Activity}
\begin{figure}[ht]
\centering
\includegraphics[width=\columnwidth,clip=0]{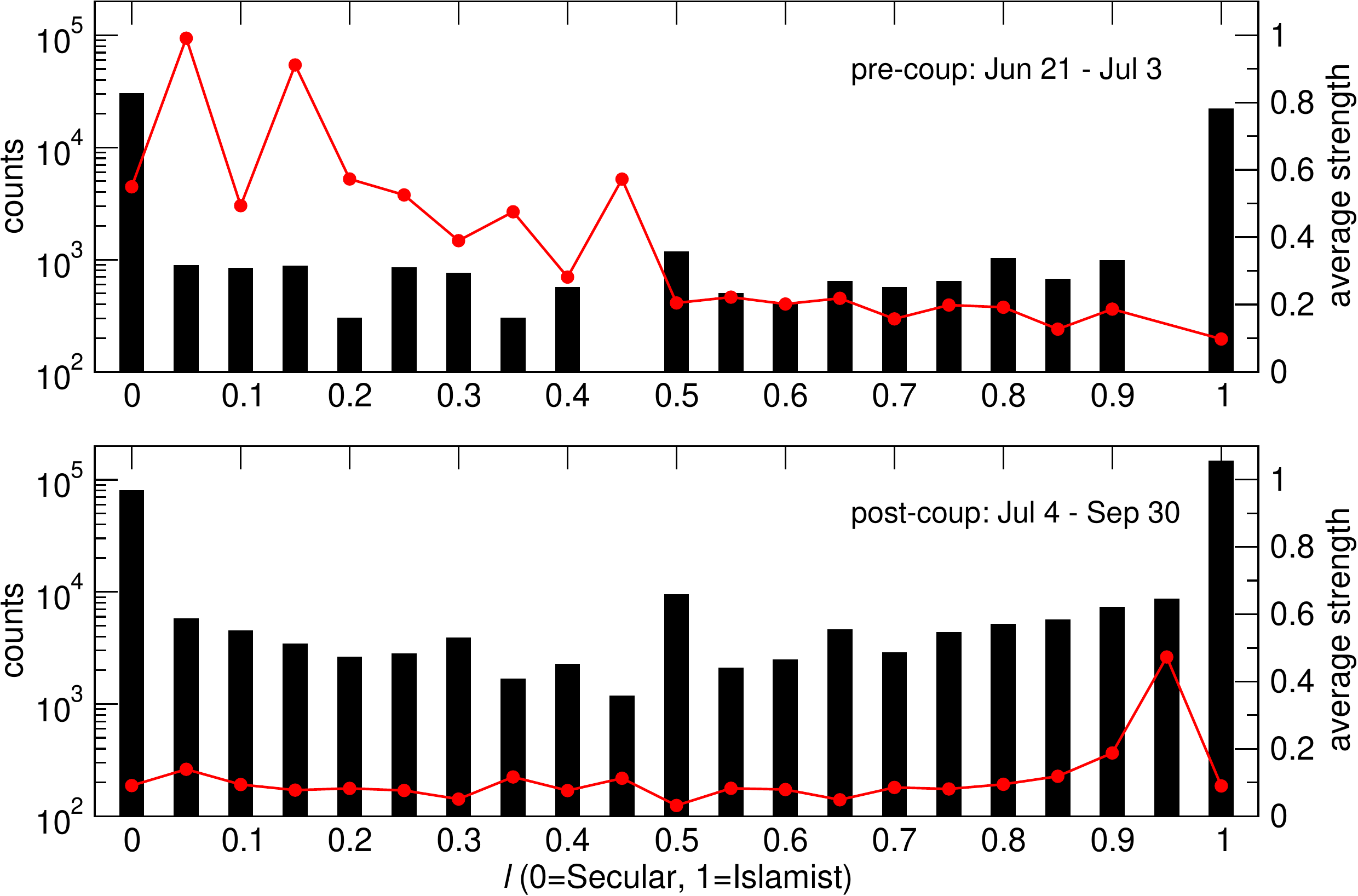}
\caption{Polarization histograms and average per-day activity (red) for two different periods (pre- and post-intervention, top and bottom panels respectively). Note logarithmic scale for histogram counts (left).}
\label{hist}
\end{figure}

Besides a longitudinal (across time) view of the leaning evolution, we aggregated all this information to provide soft labels per user. This was easily done either for the whole episode (June 21 -- September 30) or for some limited period of interest $[t_{0}, t_{f}]$. For each user, we calculated his soft label by adding up every time he was assigned an Islamist leaning and normalizing this value by the total number of times he had been present in the network over time. Thus, an $l_{i}=1$ label means that user $i$ was tagged as an Islamist for the whole period under consideration. Labels can take values in the interval $0 \le l \le 1$. In Figure~\ref{hist} we placed these soft labels in a histogram (0.05 bin width; note log scale in the histogram counts axis), to get a global view of the leaning landscape during the unfolding of events. Top panel corresponds to soft labels in the pre-coup period (June 21 -- July 3). During this first period, seculars (leaning = 0) and moderate actors ($0 < l < 0.5$) outnumber Islamist ($l = 1$) and pro-Islamic moderates ($0.5 < l < 1$); also, average activity in each of these classes is significantly higher for the former. This scenario drastically changes for the second period (bottom, July 4 -- September 30), in which Islamists recruit many new users for their side, as well as the average activity increases. On the secular side, activity drops as the size of the party decreases in relative terms.

In coherence with Figure~\ref{comsize}, the number of secular sympathizers is larger (top panel) during the pre-coup period (those with $l < 0.5$), but the opposite is observed for the post-coup period (lower panel). Furthermore, we computed the average strength (number of retweets) per user and per day within each bin (red line in each panel). Clearly, those with a secular leaning are (comparatively) much more active than Islamists during the initial period, but the activity level flattens afterwards --with a slight advantage for the Islamist side.

\section{Discussion}
At this point, we can attain a global picture of the temporal evolution of the events in Egypt during the time window of interest. On one hand, content-based analysis showed that some group \emph{pro-MI} leaning tweets were flowing in Twitter, and they were especially dominant during the brewing days before the military intervention in early July. Group \emph{anti-MI} activity became dominant soon after July 3, with increasing bursts as violent events unfold. On the other hand, political leaning in the retweet networks underwent a similar transition, i.e.\ Secularists constitute a larger and more active group up to the coup d'etat, but progressively decrease in number and activity thereafter. These two bipolar schemes (pro/anti-MI, Secular/Islamist) with similar trajectories may lead to the conclusion that Secularists have predominantly a pro-coup standpoint, while Islamist map onto an anti-coup opinion. Such was the general interpretation being projected in classic mass media during the summer of 2013 \cite{news2}.

\begin{figure}[ht]
\centering
\includegraphics[width=\columnwidth,clip=0]{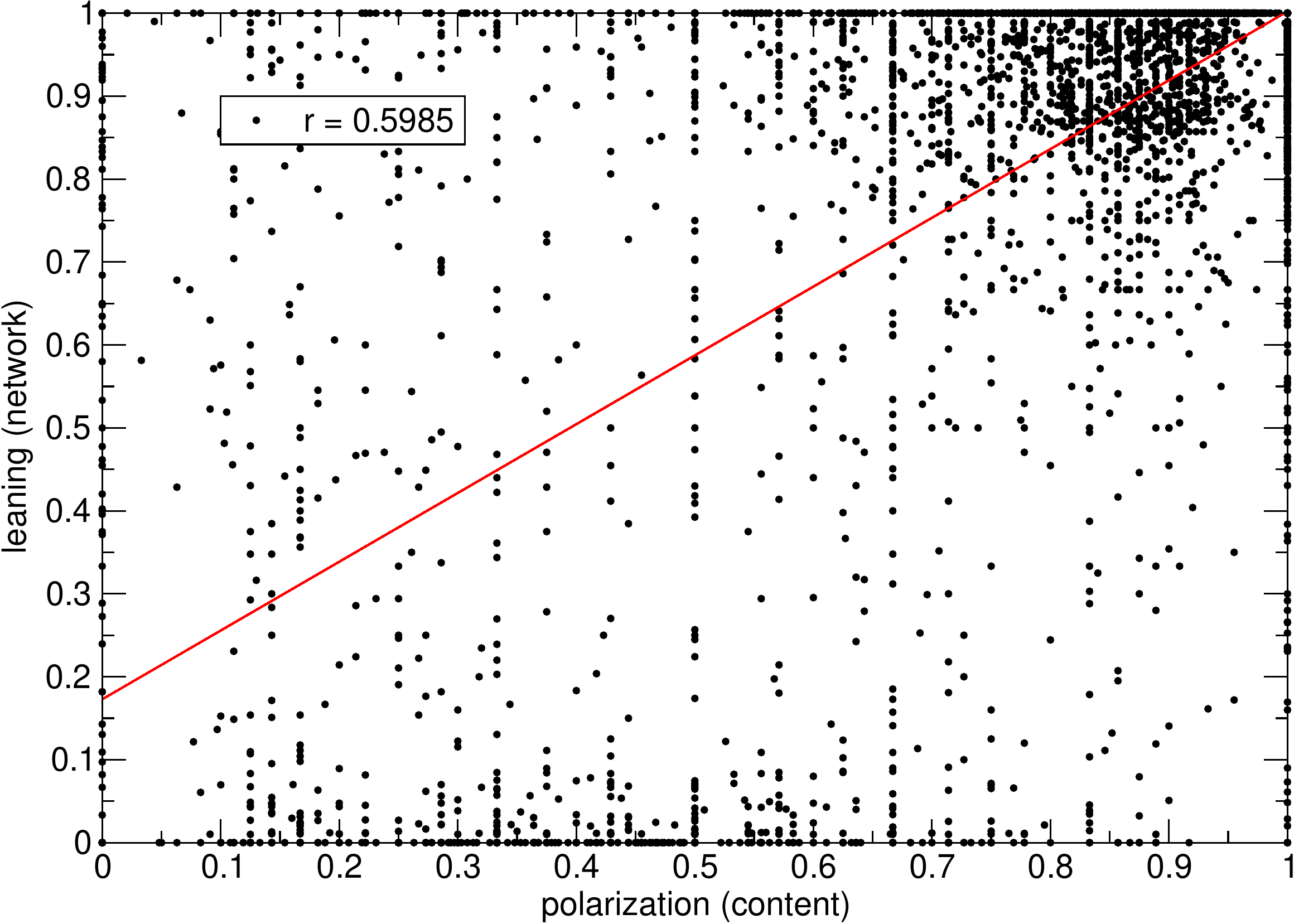}
\caption{Network-based leaning versus content-based polarity. In the x-axis, lower polarization implies a pro-MI stance, while anti-MI contents are closer to 1. In the y-axis, low leaning values correspond to the Secular camp, whereas Islamists are mapped onto higher ones.}
\label{scatter}
\end{figure}

On closer inspection, the scenario turns out to be considerably more complex. If we confront (content-based) group pro/anti-MI polarization scores and (network-based) Secularist/Islamic leaning soft labels for each Twitter users, a relatively strong correlation is found ($r \sim 0.6$, see Figure \ref{scatter}). This means that the leading trend in our analyses (which, noteworthy, have been obtained exploiting different facets of the data) exhibits a significant alignment pro-MI-to-Secularists and anti-MI-to-Islamists; and yet such alignment is far from perfect: ideally, leaning and polarity should meet around the diagonal in the scatter plot. On closer inspection, most of the disagreement occurs on the Secularist side, whereas there seems to be a rather univocal coherence in the Islamist side. Secularists exhibit rather heterogeneous positions regarding the military intervention in the country's politics.

A possible conclusion is that anti-Morsi protests were in fact quite diverse. In this direction, what we labeled here as Secularists might actually be conformed by a heterogeneous collective, including seculars indeed, as well as people without political affiliations and even small groups of Islamists --not necessarily with a pro-MI leaning. On the whole, Islamists exhibited a solid anti-MI stance as the diverse contrarians progressively withdraw. As facts have evolved during this past year, our conclusions seem to grow stronger: indeed, the provisional government led by General Sisi has prosecuted Islamist figures, but also groups or individuals from other ideological stances \cite{news3}.

Another observation that can be gleaned from our content- and network-based analyses is that switching between pro- and anti-MI groups, and between Secularists and Islamists, is rather limited. Most of the change in volume in observed polarized tweets seems to be mostly due to each group becoming louder or quieter as time progresses. This is consistent with the work of Chenoweth and Stephan~\cite{chenoweth2011civil} where they observed that members of opposing groups in conflict do not necessarily switch sides to tip the balance in favor of one of the groups, but they may merely withdraw their support.

\section{Conclusions}
In this paper we have analyzed how recent turmoil in Egypt in the period between June 21 and September 30, 2013 has transpired in Twitter.  We examine the tweets from the perspective of Secularists vs.\ Islamists and from the perspective of pro- and anti-military intervention. The main lesson from our observations is that the military takeover on July 3 caused major quantitative (volume of polarized tweets), but not ideological (polarity swaps) shifts among Twitter users. Secularist and pro-military intervention Twitter users were very loud before the take-over, but became increasingly silent afterwards. Conversely, Islamists and anti-military intervention Twitter users become significantly louder after the takeover. Furthermore, a close inspection of our results indicate that the axis Secular/Islamist can not be trivially mapped onto a pro/anti-coup opinion --which was the dominant assumption in traditional media at the time of the military intervention. This is specially true for the Secularist-to-pro-MI projection, which is not as aligned as it was generally presumed. The development of the political situation in Egypt during the past year have confirmed our insights.

To reach such conclusions we have exploited two methods, which rely on different (and non-overlapping) aspects of our dataset: the semantic and the structural dimensions. Noteworthy, the trends observed from these varied methodologies converge, on a qualitative basis.

We are aware of the limitations of the present study. Besides the inherent biases in Twitter data \cite{gonzalez2012assessing}, other arguments can be put forth. For instance, it is unlikely that an individual publishes a radical change of opinion through a single medium, i.e.\ such attitude has an associated social cost. Therefore, the availability of other sources (Facebook, blogs, etc.) would be highly desirable. We stress, however, the precise nature of the events we analyze here. We do not expect, in such bipolar scenario, that a Secularist might turn Islamist (or viceversa). But it is possible --and even {\em probable}, given the turn of the events-- that Seculars who held high expectations for a government change, were afterwards deceived as violence and repression grew. Expressing such discontent is not as costly --if it is at all.

For possible future directions, we aim to deeply analyze the motivation of users who switched polarity to understand reasons for people to change leaning. We could see some examples of users who withdraw support for the military intervention just after it occurred, others after the violent crackdown of the Rabia sit-in, and others who supported the military-intervention after opposing it in the beginning. This non-trivial switching temporal pattern resembles that of ``complex contagion'', for which a rich theoretical litearture is available \cite{borge2013cascading}, in which early adopters (early switchers, in this case) pave the way for adoption (switching) cascades. Another essential work direction is to add an additional polarization layer to the Islamist/Secular classification. From the results we have obtained, there might be a missing polarity, which could match old regime supporters (Mubarak's regime). It is known that these were heavily involved in the June 30 protests, and are the strongest supporters of the army. Accounting for them would be interesting, and can give a more complete image of the situation.

On a general framework, the work contributes to an understanding of bipolarized societies --and these are not limited to Egypt: the underlying questions and methods are more widely applicable. The turmoil in Crimea and Eastern Ukraine, the ongoing conflict in Palestine or the Catholic/Anglican dichotomy in Northern Ireland might be other promising case studies.

\bibliographystyle{acm-sigchi}

\begin{thebibliography}{}

\end{thebibliography}


\begin{thebibliography}{10}

\bibitem{news1}
These are some example news stories that illustrate how, as an example, the US administration has re-evaluated its position towards Mohamed Morsi and the Egyptian army over time:
http://www.washingtontimes.com/news/2012/jun/24/\\obama-calls-to-congratulate-morsi-an-islamist-on-w/\\ 
http://www.theguardian.com/world/2013/jul/03/egypt-obama-us-mohamed-morsi-crisis\\
http://www.theblaze.com/stories/2013/07/02/obama-calls-egypts-morsi-and-says-u-s-does-not-support-either-side\\
http://www.aljazeera.com/indepth/features\\/2013/07/2013710113522489801.html\\
http://www.forbes.com/sites/dougbandow/2014/01/20/\\
pharaoh-al-sisi-takes-control-in-egypt-obama-administration-sacrifices-security-human-rights-and-democracy

\bibitem{news2}
See for instance\\
http://www.project-syndicate.org/commentary/how-egypt-can-avoid-the-fate-of-algeria-in-1992-by--lvaro-d--vasconcelos

\bibitem{news3}
Prosecution in Egypt beyond the Muslim Brotherhood. Media and secular activists:
http://www.aljazeera.com/indepth/interactive/2014/07/\\freeajstaff-days-jailed-journalism-greste-fahmy-mohamed-b-20147166544671591.html\\
http://www.theguardian.com/world/2013/nov/27/egypt-secular-activists-arrested-protest-law\\
http://www.bbc.com/news/world-middle-east-25484064

\bibitem{adamicglance08linkkdd}
Adamic, L.~A., and Glance, N.
\newblock The political blogosphere and the 2004 u.s. election: divided they
  blog.
\newblock In {\em LinkKDD@KDD} (2005), 36--43.

\bibitem{al2012egyptian}
Al-Ani, B., Mark, G., Chung, J., and Jones, J.
\newblock The egyptian blogosphere: a counter-narrative of the revolution.
\newblock In {\em Proceedings of the ACM 2012 conference on Computer Supported
  Cooperative Work}, ACM (2012), 17--26.

\bibitem{attiaetal11ecra}
Attia, A.~M., Aziz, N., Friedman, B., and Elhusseiny, M.~F.
\newblock Commentary: The impact of social networking tools on political change
  in egypt's ``revolution 2.0''.
\newblock {\em ECRA 10\/} (2011), 369--374.

\bibitem{azab12ijep}
Azab, N.~A.
\newblock The role of the internet in shaping the political process in egypt.
\newblock {\em IJEP 3\/} (2012), 31--51.

\bibitem{borge2013cascading}
Borge-Holthoefer, J., Ba{\~n}os, R.~A., Gonz{\'a}lez-Bail{\'o}n, S., and
  Moreno, Y.
\newblock Cascading behaviour in complex socio-technical networks.
\newblock {\em Journal of Complex Networks 1}, 1 (2013), 3--24.

\bibitem{borge2011structural}
Borge-Holthoefer, J., Rivero, A., Garc{\'\i}a, I., Cauh{\'e}, E., Ferrer, A.,
  Ferrer, D., Francos, D., I{\~n}iguez, D., P{\'e}rez, M., Ruiz, G., et~al.
\newblock Structural and dynamical patterns on online social networks: the
  spanish may 15th movement as a case study.
\newblock {\em PLoS One 6}, 8 (2011), e23883.

\bibitem{chenoweth2011civil}
Chenoweth, E., and Stephan, M.~J.
\newblock {\em Why civil resistance works: The strategic logic of nonviolent
  conflict}.
\newblock Columbia University Press, 2011.

\bibitem{choudharyetal12cacm}
Choudhary, A., Hendrix, W., Lee, K., Palsetia, D., and Liao, W.-K.
\newblock Social media evolution of the egyptian revolution.
\newblock {\em CACM 55}, 5 (2012), 74--80.

\bibitem{conover2011political}
Conover, M., Ratkiewicz, J., Francisco, M., Goncalves, B., Flammini, A., and
  Menczer, F.
\newblock Political polarization on twitter.
\newblock In {\em ICWSM} (2011).

\bibitem{crammer2002algorithmic}
Crammer, K., and Singer, Y.
\newblock On the algorithmic implementation of multiclass kernel-based vector
  machines.
\newblock {\em The Journal of Machine Learning Research 2\/} (2002), 265--292.

\bibitem{igraph}
Cs\'ardi, G., and Nepusz, T.
\newblock The igraph software package for complex network research.
\newblock {\em International Journal Complex Systems\/} (2006), 1695.

\bibitem{darwish2012language}
Darwish, K., Magdy, W., and Mourad, A.
\newblock Language processing for arabic microblog retrieval.
\newblock In {\em Proceedings of the 21st ACM international conference on
  Information and knowledge management}, ACM (2012), 2427--2430.

\bibitem{magdy2014tweetmogaz}
Elsawy, E., Mokhtar, M., and Magdy, W.
\newblock Tweetmogaz v2: Identifying news stories in social media.
\newblock In {\em CIKM} (2014).

\bibitem{gayo2011limits}
Gayo-Avello, D., Metaxas, P.~T., and Mustafaraj, E.
\newblock Limits of electoral predictions using twitter.
\newblock In {\em ICWSM} (2011).

\bibitem{gonzalez2013broadcasters}
Gonz{\'a}lez-Bail{\'o}n, S., Borge-Holthoefer, J., and Moreno, Y.
\newblock Broadcasters and hidden influentials in online protest diffusion.
\newblock {\em American Behavioral Scientist 57}, 7 (2013), 943--965.

\bibitem{gonzalez2011dynamics}
Gonz\'{a}lez-Bail\'{o}n, S., Borge-Holthoefer, J., Rivero, A., and Moreno, Y.
\newblock The dynamics of protest recruitment through an online network.
\newblock {\em Scientific Reports 1\/} (2011), 197.

\bibitem{gonzalez2012assessing}
Gonz\'{a}lez-Bail\'{o}n, S., Wang, N., Rivero, A., Borge-Holthoefer, J., and
  Moreno, Y.
\newblock Assessing the bias in communication networks sampled from twitter.
\newblock {\em Social Networks 38\/} (2014), 16--27.

\bibitem{arab2011}
Governance, and Innovation~Program, D. S. o.~G.
\newblock Arab social media report, vol. 1, issue 2.
\newblock \url{www.ArabSocialMediaReport.com}, 2011.

\bibitem{jungherr2012pirate}
Jungherr, A., J{\"u}rgens, P., and Schoen, H.
\newblock Why the pirate party won the german election of 2009 or the trouble
  with predictions: A response to tumasjan, a., sprenger, to, sander, pg, \&
  welpe, im ``predicting elections with twitter: What 140 characters reveal
  about political sentiment''.
\newblock {\em Social Science Computer Review 30}, 2 (2012), 229--234.

\bibitem{khamis11ijoc}
Khamis, S.
\newblock The transformative egyptian media landscape: Changes, challenges and
  comparative perspectives.
\newblock {\em IJOC 5\/} (2011), 1159--1177.

\bibitem{lim12jc}
Lim, M.
\newblock Clicks, cabs, and coffee houses: Social media and oppositional
  movements in egypt, 2004--2011.
\newblock {\em JC 62\/} (2012), 231--248.

\bibitem{lin2013voices}
Lin, Y.-R., Margolin, D., Keegan, B., and Lazer, D.
\newblock Voices of victory: A computational focus group framework for tracking
  opinion shift in real time.
\newblock In {\em Proceedings of the 22nd international conference on World
  Wide Web}, International World Wide Web Conferences Steering Committee
  (2013), 737--748.

\bibitem{magdy2013tweetmogaz}
Magdy, W.
\newblock Tweetmogaz: a news portal of tweets.
\newblock In {\em SIGIR} (2013).

\bibitem{magdy2014adaptive}
Magdy, W., and Elsayed, T.
\newblock Adaptive method for following dynamic topics on twitter.
\newblock In {\em ICWSM} (2014).

\bibitem{metaxas2011not}
Metaxas, P.~T., Mustafaraj, E., and Gayo-Avello, D.
\newblock How (not) to predict elections.
\newblock In {\em SocialCom-PASSAT}, IEEE (2011).

\bibitem{mostak2012}
Mostak, T.
\newblock Social media as passive polling: Using twitter and online forums to
  map islamism in egypt.
\newblock Master's thesis, Center for Middle Eastern Studies Harvard
  University, 2012.

\bibitem{mustafaraj2011vocal}
Mustafaraj, E., Finn, S., Whitlock, C., and Metaxas, P.~T.
\newblock Vocal minority versus silent majority: Discovering the opionions of
  the long tail.
\newblock In {\em Privacy, security, risk and trust (passat), 2011 ieee third
  international conference on and 2011 ieee third international conference on
  social computing (socialcom)}, IEEE (2011), 103--110.

\bibitem{ohetal12icis}
Oh, O., Eom, C., and Rao, H.
\newblock Collective sense-making through the twitter service during the 2011
  egypt revolution.
\newblock In {\em ICIS} (2012).

\bibitem{poolerosenthal85ajps}
Poole, K.~T., and Rosenthal, H.
\newblock A spatial model for legislative roll call analysis.
\newblock {\em AJPS\/} (May 1985), 357--384.

\bibitem{raghavan2007near}
Raghavan, U.~N., Albert, R., and Kumara, S.
\newblock Near linear time algorithm to detect community structures in
  large-scale networks.
\newblock {\em Physical Review E 76}, 3 (2007), 036106.

\bibitem{starbirdpalen12cscw}
Starbird, K., and Palen, L.
\newblock (how) will the revolution be retweeted?: information diffusion and
  the 2011 egyptian uprising.
\newblock In {\em CSCW} (2012), 7--16.

\bibitem{venkataraman2012measuring}
Venkataraman, M., Subbalakshmi, K., and Chandramouli, R.
\newblock Measuring and quantifying the silent majority on the internet.
\newblock In {\em Sarnoff Symposium (SARNOFF), 2012 35th IEEE}, IEEE (2012),
  1--5.

\bibitem{webergarimella13asonam}
Weber, I., and Garimella, K. R.~K.
\newblock \#egypt: visualizing islamist vs. secular tension on twitter.
\newblock In {\em ASONAM} (2013), 1100--1101.

\bibitem{weberetal13asonam}
Weber, I., Garimella, V. R.~K., and Batayneh, A.
\newblock Secular vs. islamist polarization in egypt on twitter.
\newblock In {\em ASONAM} (2013), 290--297.

\bibitem{wulf2013fighting}
Wulf, V., Aal, K., Abu~Kteish, I., Atam, M., Schubert, K., Rohde, M., Yerousis,
  G.~P., and Randall, D.
\newblock Fighting against the wall: Social media use by political activists in
  a palestinian village.
\newblock In {\em Proceedings of the SIGCHI Conference on Human Factors in
  Computing Systems}, ACM (2013), 1979--1988.

\bibitem{wulf2013ground}
Wulf, V., Misaki, K., Atam, M., Randall, D., and Rohde, M.
\newblock 'on the ground'in sidi bouzid: investigating social media use during
  the tunisian revolution.
\newblock In {\em Proceedings of the 2013 conference on Computer supported
  cooperative work}, ACM (2013), 1409--1418.

\end{thebibliography}

\end{document}